\documentclass[sigplan,10pt,screen]{acmart}

\renewcommand\footnotetextcopyrightpermission[1]{}

\usepackage[utf8]{inputenc}
\usepackage[T1]{fontenc}
\usepackage[english]{babel}

\microtypecontext{spacing=nonfrench}

\usepackage{glossaries}
\glsdisablehyper
\glsdisablehyper

\def \txone {ThunderX-1\xspace}

\newacronym{acpi}{ACPI}{Advanced Configuration and Power Interface}
\newacronym{acs}{ACS}{ATA Command Set}
\newacronym{afu}{AFU}{Accelerator Function Unit}
\newacronym{ahci}{AHCI}{Advanced Host Controller Interface}
\newacronym{ascii}{ASCII}{American Standard Code for Information Interchange}
\newacronym{asic}{ASIC}{Application-Specific Integrated Circuit}
\newacronym{ata}{ATA}{AT Attachment}
\newacronym{atapi}{ATAPI}{AT Attachment Packet Interface}
\newacronym{atf}{ATF}{ARM Trusted Firmware}
\newacronym{bar}{BAR}{Base Address Register}
\newacronym{bdk}{BDK}{Board Development Kit}
\newacronym{bist}{BIST}{Built-in Self-test}
\newacronym{bmc}{BMC}{Baseboard Management Controller}
\newacronym{bram}{BRAM}{Block RAM}
\newacronym{caam}{CAAM}{Cryptographic Acceleration and Assurance Module}
\newacronym{cam}{CAM}{Common Access Method}
\newacronym{capi}{CAPI}{Coherent Accelerator Processor Interface}
\newacronym{capp}{CAPP}{Coherent Accelerator Processor Proxy}
\newacronym{cca}{CCA}{Confidential Compute Architecture}
\newacronym{ccip}{CCI-P}{Core Cache Interface}
\newacronym{ccix}{CCIX}{Cache Coherent Interconnect for Accelerators}
\newacronym{ccpi}{CCPI}{Cavium Coherent Processor Interconnect}
\newacronym{ci}{CI}{Continuous Intergration}
\newacronym{clut}{CLUT}{Cache Line Under Test}
\newacronym{cpld}{CPLD}{Complex Programmable Logic Device}
\newacronym{cpu}{CPU}{Central Processing Unit}
\newacronym{cxl}{CXL}{Compute eXpress Link}
\newacronym{dac}{DAC}{Digital Analog Converter}
\newacronym{dc}{DC}{Display Controller}
\newacronym{dep}{DEP}{Data Execution Prevention}
\newacronym{dma}{DMA}{Direct Memory Access}
\newacronym{dos}{DoS}{denial-of-service}
\newacronym{dsp}{DSP}{Digital Signal Processor}
\newacronym{eci}{ECI}{Enzian Coherence Interface}
\newacronym{edma}{eDMA}{Enhanced Direct Memory Access}
\newacronym{ept}{EPT}{Extended Page Table}
\newacronym{esai}{ESAI}{Enhanced Synchronous Audio Interface}
\newacronym{esp}{ESP}{Executable Space Protection}
\newacronym{fat}{FAT}{File Allocation Table}
\newacronym{fis}{FIS}{Frame Information Structure}
\newacronym{fiu}{FUI}{FPGA Interface Unit}
\newacronym{fmc}{FMC}{FPGA Mezzanine Card}
\newacronym{fpga}{FPGA}{Field-Programmable Gate Array}
\newacronym{fsis}{FSIS}{Filesystem Information Sector}
\newacronym{fsm}{FSM}{Finite State Machine}
\newacronym{gpu}{GPU}{Graphics Processing Unit}
\newacronym{gqr}{GQR}{Generalized-Quasi-Reduction}
\newacronym{gsync}{\textsc{GSync}}{Global Synchronization}
\newacronym{hba}{HBA}{Host Bus Adapter}
\newacronym{hbm}{HBM}{High-Bandwidth Memory}
\newacronym{hpc}{HPC}{High-Performance Computing}
\newacronym{i2c}{I\textsuperscript{2}C}{Inter-Integrated Circuit}
\newacronym{ich}{ICH}{I/O Controller Hub}
\newacronym{ic}{IC}{Integrated Circuit}
\newacronym{idc}{IDC}{Inter-Domain Communication}
\newacronym{iee}{IEE}{Inline Encryption Engine}
\newacronym{ipc}{IPC}{Inter-Process Communication}
\newacronym{ipi}{IPI}{Inter-Processor Interrupt}
\newacronym{ip}{IP}{intellectual property block}
\newacronym{ipmi}{IPMI}{Intelligent Platform Management Interface}
\newacronym{iut}{IUT}{Implementation Under Test}
\newacronym{kaslr}{KASLR}{Kernel Address Space Layout Randomization}
\newacronym{lba}{LBA}{Logical Block Address}
\newacronym{llc}{LLC}{Last-Level Cache}
\newacronym{lru}{LRU}{Least-recently used}
\newacronym{mmio}{MMIO}{Memory-Mapped I/O}
\newacronym{mmu}{MMU}{Memory Management Unit}
\newacronym{mpsoc}{MPSoC}{Multiprocessor System-on-a-Chip}
\newacronym{mpu}{MPU}{Memory Protection Unit}
\newacronym{mpx}{MPX}{Memory Protection Extensions}
\newacronym{msi}{MSI}{Message-Signalled Interrupt}
\newacronym{mu}{MU}{Messaging Unit}
\newacronym{ndp}{NDP}{Near-Data Processing}
\newacronym{ncq}{NCQ}{Native Command Queueing}
\newacronym{nic}{NIC}{Network Interface Adaptor}
\newacronym{nmp}{NMP}{Near-Memory Processing}
\newacronym{numa}{NUMA}{Non-Uniform Memory Access}
\newacronym{nvme}{NVMe}{NVM Express}
\newacronym{ocapi}{OpenCAPI}{Open Coherent Accelerator Processor Interface}
\newacronym{oci}{OCI}{Octeon III multi-node Interconnect}
\newacronym{os}{OS}{Operating System}
\newacronym{pae}{PAE}{Physical Address Extensions}
\newacronym{pata}{PATA}{Parallel ATA}
\newacronym{pcb}{PCB}{Printed Circuit Board}
\newacronym{pcie}{PCIe}{PCI Express}
\newacronym{pci}{PCI}{Peripheral Component Interconnect}
\newacronym{piix}{PIIX}{PCI IDE ISA Xcelerator}
\newacronym{pim}{PIM}{Processing In Memory}
\newacronym{pio}{PIO}{Programmed I/O}
\newacronym{pmbus}{PMBus}{Power Management Bus}
\newacronym{prd}{PRD}{Physical Region Descriptor}
\newacronym{psci}{PSCI}{Power State Coordination Interface}
\newacronym{psl}{PSL}{POWER Service Layer}
\newacronym[longplural={Page Table Entries}]{pte}{PTE}{Page Table Entry}
\newacronym{qpi}{QPI}{QuickPath Interconnect}
\newacronym{rdma}{RDMA}{Remote Direct Memory Access}
\newacronym{rfis}{RFIS}{Received FIS}
\newacronym{rpc}{RPC}{Remote Procedure Call}
\newacronym{rtc}{RTC}{Real Time Clock}
\newacronym{sai}{SAI}{Synchronous Audio Interface}
\newacronym{sata}{SATA}{Serial ATA}
\newacronym{sbc}{SBC}{single board computer}
\newacronym{scsi}{SCSI}{Small Computer System Interface}
\newacronym{scu}{SCU}{System Controller Unit}
\newacronym{sdk}{SDK}{Software Developement Kit}
\newacronym{seco}{SECO}{Security Controller}
\newacronym{sgx}{SGX}{Software Guard Extensions}
\newacronym{simd}{SIMD}{Single Input Multiple Data}
\newacronym{sim}{SIM}{SCSI Interface Module}
\newacronym{skb}{SKB}{System Knowledgebase}
\newacronym{smbus}{SMBus}{System Management Bus}
\newacronym{smc}{SMC}{Secure Monitor Call}
\newacronym{smm}{SMM}{System Management Mode}
\newacronym{smmu}{SMMU}{System Memory Management Unit}
\newacronym[longplural={Systems-on-Chip}]{soc}{SoC}{System-on-Chip}
\newacronym[longplural={Systems-on-Module}]{som}{SoM}{System-on-Module}
\newacronym{spl}{SPL}{System Protocol Layer}
\newacronym{tap}{TAP}{Test Access Port}
\newacronym{tcb}{TCB}{Trusted Computing Base}
\newacronym{tcm}{TCM}{Tightly Coupled Memory}
\newacronym{tcp}{TCP}{Transmission Control Protocol}
\newacronym{tdp}{TDP}{Thermal Design Power}
\newacronym{tee}{TEE}{Trusted Execution Environment}
\newacronym{tfa}{TF-A}{Trusted Firmware-A}
\newacronym{tlb}{TLB}{Translation Lookaside Buffer}
\newacronym{tpu}{TPU}{Tensor Processing Unit}
\newacronym{ttbr}{TTBR}{Translation Table Base Register}
\newacronym{uart}{UART}{universal asynchronous receiver-transmitter}
\newacronym{uefi}{UEFI}{Unified Extensible Firmware Interface}
\newacronym{upi}{UPI}{Universal Path Interconnect}
\newacronym{usb}{USB}{Universal Serial BUS}
\newacronym{vfpga}{vFPGA}{Virtual FPGA}
\newacronym{vfs}{VFS}{Virtual Filesystem}
\newacronym{vliw}{VLIW}{Very Long Instruction Word}
\newacronym{vpu}{VPU}{Video Processing Unit}
\newacronym{xmpu}{XMPU}{Xilinx Memory Protection Unit}
\newacronym{xppu}{XPPU}{Xilinx Peripheral Protection Unit}
\newacronym{xrdc}{XRDC}{Extended Resource Domain Controller}
\newacronym{thp}{THP}{Transparent Huge Page}
\newacronym{ddio}{DDIO}{Direct Data I/O}
\newacronym{dbms}{DBMS}{Database Management System}
\newacronym{gc}{GC}{Garbage Collection}

\newacronym{amac}{AMAC}{Asynchronous Memory Access Chaining}
\newacronym{dcs}{DCS}{Directory Controller Slice}
\newacronym{mig}{MIG}{Memory Interface Generator}
\newacronym{csr}{CSR}{compressed sparse row}
\newacronym{clfr}{CLFR}{cache line fetch rate}

\glsunset{os}
\glsunset{fpga}
\glsunset{pci}
\glsunset{pcie}
\glsunset{usb}

\usepackage{multirow}

\newcommand{\ndp}{\gls{ndp}\xspace}
\newacronym{cp}{CP}{channel program}
\newacronym[longplural=channel binaries]{cb}{CB}{channel binary}
\newacronym{mcc}{MCC}{memory channel controller}
\newacronym{uts}{UTS}{Unbalanced Tree Search}
\newcommand{\structera}{Structera-A\xspace}
\newcommand{\upmem}{UPMEM\xspace}

\RequirePackage{listings}
\newcommand*\code[1]{\lstinline[basicstyle=\ttfamily]{#1}}

\usepackage[ruled,vlined,linesnumbered]{algorithm2e}
\RequirePackage{algpseudocode}

\colorlet{colorcpu}{orange!20}
\colorlet{colormcc}{blue!20}

\SetCommentSty{mycommfont}

\usepackage{xspace}
\newcommand{\system}{Proxics\xspace}

\newcommand{\mcc}{\gls{mcc}\xspace}
\newcommand{\mccs}{\glspl{mcc}\xspace}
\newcommand{\cp}{\gls{cp}\xspace}
\newcommand{\cps}{\glspl{cp}\xspace}

\newcommand{\MMNDP}{M\textsuperscript{2}NDP\xspace}

\usepackage{xcolor}
\usepackage{todonotes}

\usepackage{subcaption}

\makeatletter
\let\old@lstKV@SwitchCases\lstKV@SwitchCases
\def\lstKV@SwitchCases#1#2#3{}
\makeatother
\usepackage{lstlinebgrd}
\makeatletter
\let\lstKV@SwitchCases\old@lstKV@SwitchCases

\lst@Key{numbers}{none}{%
    \def\lst@PlaceNumber{\lst@linebgrd}%
    \lstKV@SwitchCases{#1}%
    {none:\\%
     left:\def\lst@PlaceNumber{\llap{\normalfont
                \lst@numberstyle{\thelstnumber}\kern\lst@numbersep}\lst@linebgrd}\\%
     right:\def\lst@PlaceNumber{\rlap{\normalfont
                \kern\linewidth \kern\lst@numbersep
                \lst@numberstyle{\thelstnumber}}\lst@linebgrd}%
    }{\PackageError{Listings}{Numbers #1 unknown}\@ehc}}
\makeatother

\renewcommand\footnotetextcopyrightpermission[1]{}
\newcommand*\circled[1]{\raisebox{.5pt}{\textcircled{\raisebox{-.9pt} {#1}}}}

\begin{document}

\title[{\system: an efficient programming model for far memory accelerators}]{\system: an efficient programming model\\ for far memory accelerators}

\author{Zikai Liu}
\email{zikai.liu@inf.ethz.ch}
\affiliation{%
    \institution{ETH Zurich}
    \city{Zurich}
    \country{Switzerland}
}
\orcid{0009-0000-5411-9785}

\author{Niels Pressel}
\email{npressel@ethz.ch}
\affiliation{%
    \institution{ETH Zurich}
    \city{Zurich}
    \country{Switzerland}
}
\orcid{0009-0006-3544-9037}

\author{Jasmin Schult}
\email{jasmin.schult@inf.ethz.ch}
\affiliation{%
    \institution{ETH Zurich}
    \city{Zurich}
    \country{Switzerland}
}
\orcid{0009-0000-1815-3206}

\author{Roman Meier}
\email{roman.meier@inf.ethz.ch}
\affiliation{%
    \institution{ETH Zurich}
    \city{Zurich}
    \country{Switzerland}
}
\orcid{0009-0001-9165-0046}

\author{Pengcheng Xu}
\email{pengcheng.xu@inf.ethz.ch}
\affiliation{%
    \institution{ETH Zurich}
    \city{Zurich}
    \country{Switzerland}
}
\orcid{0000-0002-2724-7893}

\author{Timothy Roscoe}
\email{troscoe@inf.ethz.ch}
\affiliation{%
    \institution{ETH Zurich}
    \city{Zurich}
    \country{Switzerland}
}
\orcid{0000-0002-8298-1126}

\begin{abstract}
The use of \emph{disaggregated} or \emph{far memory} systems such as CXL
memory pools has renewed interest in Near-Data Processing (NDP):
situating cores close to memory to reduce bandwidth requirements to
and from the CPU.  Hardware designs for such accelerators are
appearing, but there lack clean, portable OS abstractions
for programming them.

We propose a programming model for NDP devices based on
familiar OS abstractions: virtual processors (processes) and
inter-process communication channels (like Unix pipes). 

While appealing from a user perspective, a naive \emph{implementation} of
such abstractions is inappropriate for NDP accelerators: the paucity
of processing power in some hardware designs makes classical processes
overly heavyweight, and IPC based on shared buffers makes no sense in
a system designed to reduce memory bandwidth.

Accordingly, we show how to implement these abstractions in a
lightweight and efficient manner by exploiting compilation 
and interconnect protocols.  We demonstrate them with a real
hardware platform runing applications with a range of memory access
patterns, including bulk memory operations, in-memory databases and
graph applications.

Crucially, we show not only the benefits over CPU-only
implementations, but also the critical importance of \emph{efficient,
low-latency communication channels} between CPU and NDP accelerators,
a feature largely neglected in existing proposals.

\end{abstract}

\maketitle
\pagestyle{plain}

\section{Introduction} \label{sec:intro}

\emph{Far memory} -- DRAM not directly connected to a CPU but instead
accessed at a cache-line granularity over an interconnect fabric -- is
beginning to be deployed widely in
data centers~\cite{Elyse:AzureCXL:2025}, for example using
\gls{cxl}~\cite{CXL31}.

Far memory expansion has several attractive features: far RAM can
be \emph{disaggregated} and allocated dynamically to different
servers; it allows reuse of older DRAM modules at a time when DIMM
prices are rising; it can even be used to share data in main memory
between machines~\cite{Berger:Octopus:2025}.

However, far memory has both higher latency and lower bandwidth than a
server's native
RAM~\cite{Li:Pond:2023,Liu:CXLPerf:2025,Sun:2023:DemystifyingCXLMemory},
raising the question of how to efficiently make use of it in
applications.  

One approach is to use memory tiering, a form of demand paging which
dynamically migrates frequently used pages into local memory and
moves those less frequently referenced further away.
While this requires relatively few changes to the OS to
manage~\cite{Zhong:2024:ManagingMemoryTiers}, it relies on the OS
and/or hardware ``second-guessing'' forthcoming access patterns which might, in practice, be known to the application.

For this reason, \gls{ndp}, whereby the OS and
applications can move computation to processor cores close to memory,
is seeing a revival, with the recent
announcement of dedicated far memory accelerators like Marvell's
\structera\cite{Marvell:StructeraA:2024}.

Such hardware comes with vendor-specific low-level SDKs, but
the art of programming them efficiently is in its infancy.
Moreover, there are no portable OS abstractions for such devices which
match the benefits they might provide~\cite{Barbalace:NDP-OS:2017}.

\textbf{\system} is a programming model which addresses this problem, allowing
developers to use common abstractions like
processes and pipes to move application functions to far-memory
accelerators more efficiently than with GPU-oriented models like
CUDA~\cite{NVIDIA:CUDA:2026} or pTasks~\cite{rossbach_ptask_2011}.
The abstractions it provides can be implemented efficiently
on far-memory accelerators: we base our evaluation on a complete,
functional implementation using research hardware.

\system not only provides a clean way to offload \emph{computation} to cores
close to far memory, but also provides efficient and fine-grained \emph{communication}
between these cores and the host CPU.  In doing so, it greatly extends
to set of applications which benefit from far memory acceleration, and
also provides insights for future hardware designs.

In the next section we provide more background on far memory
accelerators and motivating applications, before presenting the
\system model in \autoref{sec:design}. In \autoref{sec:impl}
we describe our implementation in detail, and characterize its
performance relative to announced products.  In \autoref{sec:eval} we
then show several use cases which benefit from acceleration, and
exercise increasing features of \system, before describing
implications and further questions in \autoref{sec:discussion}.

\system, together with all the application use-cases discussed in this
paper, is publicly available open source.\footnote{We will provide
the repository URL in the final version of the paper.}

\section{Background and related work} \label{sec:bg}

CXL-based far memory in the form of memory
expanders~\cite{Marvell:StructeraX:2024,Micron:CXLExpansion:2023,Samsung:CXLExpansion:2022,MontageTech:CXLExpander:2026}
and memory
pools~\cite{AsteraLabs:CXLController:2025,Samsung:CMM-B:2024,Li:Pond:2023,Berger:Octopus:2025}
is being deployed in large datacenters.  Such deployments allow
elastic scaling in memory by \emph{disaggregating} a portion of a
server's memory, allowing it to be reallocated to another server.
Far memory also promises extended capacity and lower hardware costs
by, among other things, simplifying board design and allowing reuse of
DRAM in new
machines~\cite{Li:Pond:2023,Berger:Octopus:2025,Ahn:CXL-SAP-HANA:2024,Liu:CXLPerf:2025,Ewais:DisaggregatedMemDC:2023,Korolija:Farview:2021,Maruf:TPP:2023,Elyse:AzureCXL:2025}.

Almost all existing far memory devices use the CXL 1.1 and
2.0 standards.  Measurements and published performance figures for
these devices show, unsurprisingly, significantly increased latency and reduced
bandwidth between the CPU and far
memory.
Evaluations~\cite{Wang:CXLPerf:2025,Sun:2023:DemystifyingCXLMemory,Tang:CXLPerf:2024,WangXi:PerfCXL:2025,Berger:Octopus:2025,
  Li:Pond:2023,Liu:CXLPerf:2025} report off-chip bandwidth to the CPU of 20-80
GB/s.  In contrast, the DRAM controllers of CXL controllers can typically deliver more
than 100 GB/s \emph{on-chip} bandwidth, while Marvell's Structera-A part
claims 200 GB/s~\cite{Marvell:StructeraA:2024,GoogleCloudC4A:2026}. 

This performance gap has been
attributed to the inefficiency of CXL
controllers~\cite{Wang:CXLPerf:2025,Sun:2023:DemystifyingCXLMemory,Tang:CXLPerf:2024},
but scaling these devices incurs significant real-world cost~\cite{Berger:Octopus:2025}. 

Similarly, the internal latency within CXL controllers is several
times lower than external latency to the CPU, which is between
200-400ns on current
hardware~\cite{Wang:CXLPerf:2025,Sun:2023:DemystifyingCXLMemory,Tang:CXLPerf:2024,WangXi:PerfCXL:2025,Berger:Octopus:2025,Li:Pond:2023,Liu:CXLPerf:2025}. For
comparision, accessing CPU-local DRAM typically takes around 100 ns.

\subsection{Approaches to integrating far memory}

Faced with this disparity in performance between local and far memory,
two main approaches have been proposed.  The first is tiering:
dynamically migrating pages between CPU-local DRAM and far memory,
either explicitly in the OS using variants on demand paging, or
transparently in hardware in the case of Intel's ``flat memory''
mode~\cite{Maruf:TPP:2023,Xu:FlexMem:2024,Zhong:2024:ManagingMemoryTiers}. 

This approach has the benefit of requiring no new abstractions in the
OS or application: the complex memory hierarchy can be treated by
application software as essentially flat.  However, the downside is
that performance of applications depends critically on how the access
pattern interacts with the page replacement policy.  Experience with
similar effects involving the cache hierarchy, or TLB coverage,
suggests that, sooner or later, programmers will have to optimize
their code for this new, complex memory model.

The second approach, therefore, applies the idea of \gls{ndp} to
far memory, moving computation to cores closer to remote
DRAM, exploiting the relatively high bandwidth and low latency inside
the far memory controller to reduce data that moves to and
from the CPU.  A range of hardware devices have been proposed to
implement this~\cite{HokyoonLee:CMM-DC:2025,Lee:CXL-PIM:2025,Ham:GeneralNDP:2024,Sim:CMS:2022,Huangfu:Beacon:2022,Jan:CXLANNS:2023}.
For instance, Marvell's \structera has a cluster of Arm NeoverseV2
cores closely coupled to CXL-attached
memory~\cite{Marvell:StructeraA:2024}. 

\gls{ndp} differs from \gls{pim} systems, which deeply
integrate small cores with DRAM
DIMMs~\cite{Gomez-Luna:PrIM:2022,Teguia:vPIM:2024a}, minimizing 
data movement but imposing challenging constraints on
placement, computation, and the ability of the OS to virtualize
them effectively. 
In contrast, \gls{ndp} systems are much more flexible, resembling more
conventional cores with access to a wide memory space.  However,
existing \gls{ndp} software support lacks traditional OS-oriented features:
clean and portable abstractions, memory protection, isolation,
coordination with the host CPU,
etc.~\cite{Gao:NDPAnalytics:2015,Ghose:PIMAdoption:2019,Barbalace:NDP-OS:2017,Liu:APSysMCC:2025}.

\subsection{Applications for far memory acceleration} \label{subsec:apps}

Exposing \gls{ndp} to applications or the OS has a number of potential benefits.
At the simplest level are the ``applications'' also motivating
\gls{pim}: \textbf{bulk memory operations} like memory
zeroing~\cite{Kamath:MC2:2024,Yang:Zeroing:2011,Panwar:HawkEye:2019},
copy-on-write~\cite{Kamath:MC2:2024}, hypervisor
paging~\cite{Kwon:Ingens:2017,Pham:LargePagesVM:2015}, and VM
migration~\cite{Clark:VMLiveMigration:2005,Choudhary:SurveyVMMigration:2017}.

More complex applications benefit from moving operators that reduce
data volumes closer to far memory, in particular \textbf{pushdown of
  selection and aggregation} operators in
databases~\cite{Korolija:Farview:2021,Lerner:CXLScaleUpDB:2024,Ahn:CXL-SAP-HANA:2024,Giles:ACIDCXL:2025,Woods:Ibex:2014,Jo:YourSQL:2016}
to reduce data movement.

Workloads with irregular access patterns, in particular many \textbf{graph
processing} use
cases~\cite{Lumsdaine:Challenges:2007,Zhang:Polymer:2015}, can benefit
from moving the pointer-chasing logic closer to slower memory.

Existing efforts to mitigate memory latency in these applications
often try to sustaining a large number of in-flight memory accesses
using
coroutines~\cite{Nelson:GrappaATC:2015,Nelson:GrappaTR:2014,Li:GastCoCo:2024,Zhi:CoroGraph:2023},
a technique also used in
databases~\cite{Psaropoulos:InterleavingCoro:2017,He:CoroBase:2020},
in particular \gls{amac}~\cite{Kocberber:AMAC:2015}.

In these cases, the benefit of existing \gls{ndp} systems is limited
because relatively fine-grained coordination between memory accesses
and the main CPU (which performs the per-node processing) is still
required.  If communication between CPU and \gls{ndp} accelerator
incurs significant overhead, the latter is reduced to essentially a
prefetcher.
This also applies to communication \emph{between} \gls{ndp} cores: for
example, Pregel~\cite{Malewicz:Pregel:2010} and
Giraph~\cite{Ching:Giraph:2015} both use fast message passing to
parallelize computation.

Finally, \textbf{PageRank}~\cite{Page:PageRank:1999} has both a
regular access pattern (graph traversal) and an irregular pattern
(score propagation).  In practice, the latter dominates run
time and is suitable to be offload to the near-memory processors,
while the former benefits from the strong compute power of the CPU.

\section{\system abstractions} \label{sec:design}

\begin{figure}[tbp]  %
    \centering
    \includegraphics[width=0.9\columnwidth]{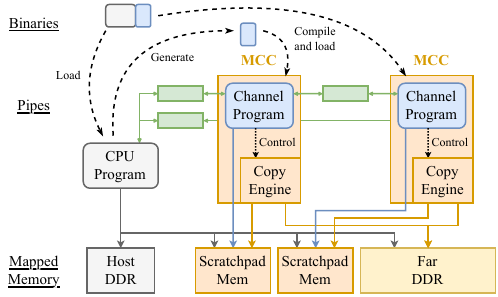}
    \caption{\system abstractions}
    \label{fig:abstraction}
\end{figure}

A common feature of current \gls{ndp} hardware is nature of software
support, which comes in the form a low-level SDK for running software
on the accelerator, without a clear corresponding OS abstraction~\cite{Barbalace:NDP-OS:2017}.
Each device is therefore programmed differently, and with few reusable
concepts other than vendor-specific low-level programming. 

In contrast, \system provides abstractions which are both
portable and efficient.  More specifically, our requirements in
designing \system were as follows:

Firstly, the model must be \textbf{efficient}: 
applications should benefit from performance close to that
achievable with hand-coded near-memory accelerator code running ``on
the metal''.  This efficiency must also be achievable at a fine
granularity, since \system targets accelerators that are trying to
reduce the overhead of memory accesses, rather than improving
coarse-grain computational throughput.
\system achieves this by providing quite a low-level abstraction of
the hardware, for example making memory operation split-phase, as
described below. 

Secondly, \system must be a \textbf{flexible} model, able
to accommodate a range of different applications' use-cases and access
patterns and deliver benefit (where possible) to all of them. 
This requirement led us to pay particular attention in \system to
fine-grained, low-latency communication, both between host CPU and
\gls{ndp} accelerators and \emph{among the accelerator cores themselves}.

Thirdly, \system's abstractions should not overly restrict the hardware
architectures on which it runs; in other words, it must be possible to
map the computational model onto a range of emerging accelerators.
However, a ``lowest common denominator'' approach to portability is
strongly at odds with the requirement for efficiency, and a strategic
goal of developing \system is to inform future hardware designs.  This
shows again in the communication architecture of \system.  While it
can be implemented straightforwardly on current hardware designs, when
\system can take advantage of particularly efficient communication
between processes (as we show in \autoref{subsec:eval:uts} and
\autoref{subsec:pagerank}) it enables benefits for a greatly
extended range of applications.

Finally, \system should present a model that makes intuitive sense in
the context of regular OS programming.

These requirements, and the inevitable tensions between them, led us
to adopt abstractions which might, at first sight, be surprising:
processes and pipes.

\subsection{Channel programs and processes}

In \system, application processes on the host CPU create and interact
with analogous processes on the \gls{ndp} accelerators.  Each physical
core on an accelerator is referred to as a \gls{mcc}, by analogy with
the IBM mainframe channel controllers that perform operations close to
disk storage.  \glspl{mcc} are abstracted in the host OS as
\emph{single-threaded} \gls{cp} process, which runs an
application-supplied \gls{cb} to completion.

Processes make intuitive sense as an \emph{abstraction}: they are
portable across architectures, and the code running within them can be
optimized at compile time to provide an efficient mapping to a given
hardware platform.

However, a classical Unix-like process \emph{implementation} are not a
good match for \gls{ndp} accelerators: they are heavyweight, and not
differentiated based on which cores they run on.
For this reason, \system \glspl{cp} are created using a
\texttt{spawn()} operation rather than
\texttt{fork()}~\cite{Baumann:Fork:2019}: a forking model makes no
sense when the accelerator might have a different instruction set
architecture, might not fully share memory with the parent process on
the host CPU, and might not have sufficient resources to execute a
complete copy of the parent.

Moreover, a \texttt{spawn()} model allows the child program to be
specified explicitly, which allows \system \glspl{cp} to be written in
a \emph{high-level portable form} and compiled at runtime into a
highly efficient, compact form -- indeed, multiple \glspl{cp} can even
be safely cooperatively scheduled on a single \gls{mcc}.  The
\glspl{cp} we evaluate in \autoref{sec:eval} require between
1KiB and 16KiB bytes of compiled program text.

\subsection{The \gls{cp} execution model} \label{subsec:cp-exec-model}

The execution environment for a \gls{cp} aims to maximize
flexibility and efficiency by providing a relatively low-level
abstraction which can be compiled to native code for a wide range of
hardware platforms.  It consists of one single-threaded processor
core, some scratchpad memory, a split-phase copy unit to move data to
and from memory local to the \gls{cp}, and efficient pipe-like
channels for communication with the CPU and other \glspl{cp}
serving the same application.

Few assumptions are made about the features of \gls{mcc} core.  In
particular, it is not expected to have a cache, but instead it is
assumed to be able to access its scratchpad memory with 1 or 2 cycles
of latency - similar to an L1 cache on a conventional processor.  This
provides portability: for specialized platforms like \upmem, or the
prototype use in this paper, it corresponds quite well to the
hardware.  For more powerful \gls{ndp} processors like \structera, the
processor's cache hierarchy provides an approximation to this model.

While a scratchpad differs from the traditional cache hierarchy for
CPUs, it is common in embedded systems and recent GPUs also
allow retargeting part of the L1 cache as fast scratchpad~\cite{Krashinsky:NVAmpereArch:2020}.

The copy unit moves data between scratchpad memory and attached DRAM
in larger units, corresponding to a DRAM row or a host CPU cache line
(128 bytes in our prototype, for example).  This also provides
portability by modeling \gls{ndp} accelerators with direct access to
DRAM memory channels, as in our prototype.

Importantly, the copy unit abstraction has a \emph{split-phase}
interface: a transfer between scratchpad memory and DRAM is initiated
by the \gls{cp} which can later poll for completion.  On platforms
with highly deterministic core execution speed (due to the lack of
cache) and closely-coupled DRAM channels this allows the compiler to
optimize the \gls{cp} binary offline, overlapping multiple in-flight
memory operations with computation as in the Beehive
processor~\cite{thacker_beehive}.
This makes it very natural to implement techniques like
\gls{amac}~\cite{Kocberber:AMAC:2015}, variants of which are already
widely used in optimizing database and graph computations. 

The split-phase nature of the copy unit can be emulated
straightforwardly on hardware that doesn't provide it, though
potentially at some cost in performance: superscalar processors will
transparently allow cache fills and write-backs to overlap with subsequent
(non-dependent) instructions, for example.  However, providing an
apparently-synchronous interface to DRAM using an in-order processor
would result in significant loss in efficiency. 

The copy unit can also move data between the host CPU's main memory
and scratchpad, though in practice this function is often subsumed by
the communication channels.

\subsection{Communication with pipes}
\label{subsec:communication-with-pipes}

The final part of the programming model is concerned with
communication between \glspl{cp} and the host CPU.  \system provides
asynchronous channels both between each
\gls{cp} and the host CPU, and pairwise between all \glspl{cp} within
an application.  In the abstract, these channels resemble Unix pipes:
data is written into the pipe by one end-point and read out by the
other.

Classical Unix pipes are a somewhat heavyweight primitive for
\gls{ipc} since they copy data into and out of the kernel's buffer
cache, held in main memory.  For that reason, pipes might seem an
unusual choice for lightweight communication in accelerators whose
goal is to \emph{reduce} memory access overhead.

However, as with processes, pipes admit a much more lightweight and
efficient implementation when communicating between devices.  Hardware
FIFOs between \glspl{mcc} are straightforward to implement, and while
the pipe abstraction could be provided to user-space host code via
descriptor-based DMA from an \gls{ndp} accelerator, as we show in our
implementation, the abstraction allows us to adopt much lower-latency
solutions be exploiting programmed I/O to PCIe and CXL devices, and
even exploit the CPU's cache coherent protocol.

Pipes can therefore support both streaming of data at a range of
granularities, and very low latency synchronization between processes. 
Crucially, as we show in \autoref{subsec:eval:uts}, this allows a much
broader range of applications to be benefit from \gls{ndp} than is
possible with a more coarse-grained, GPU-style model of communication.

\subsection{Discussion}

In contrast with GPU-oriented programming models such as
CUDA~\cite{NVIDIA:CUDA:2026} and pTasks~\cite{rossbach_ptask_2011}, \system
targets workloads that can be highly irregular. The model of dataflow is not
ideal for many of these applications. 
Furthermore, the \system programming model is not computation-oriented but
rather to move the computation and mitigate the off-chip memory bottleneck.
Indeed, the \mcc core are general-purpose cores rather than GPU cores or other
customized computing elements, distinct from the SIMD cores in
\MMNDP~\cite{Ham:GeneralNDP:2024}, the FPGA IPs in
FarView~\cite{Korolija:Farview:2021} and the previous \gls{pim}
devices~\cite{Gomez-Luna:PrIM:2022,Teguia:vPIM:2024a,Lee:CXL-PIM:2025}.

In addition, the use of general-purpose cores enables flexible multiplexing and
scheduling in \system. Later shown in \autoref{subsec:eval:reprogram}, \mcc can
be swiftly reprogrammed with a different \cp. \autoref{subsec:pagerank} further
shows that multiple CPU cores and \mcc cores can be arbitrarily combined. Built
on top, time-sharing multiplexing can be built as future
work(\autoref{subsec:future-os-support}). 
In contrast, while there are attempts
to time-share
GPUs~\cite{Coppock:LithOS:2025,Han:GPUPreemption:2022,Ng:Paella:2023},
FPGAs~\cite{Korolija:Coyote:2020,Ramhorst:CoyoteV2:2025,Li:Harmonia:2025,Putnam:Catapult:2014}
and \gls{pim} devices~\cite{Teguia:vPIM:2024a,Barbalace:NDP-OS:2017}, they are
not as flexible and have heavier overhead.

In this work, we present a relatively low-level programming model of \system.
A higher-level abstraction can further improve the usability of the system, but
would most likely be more specialized and could be layered on top. Examples include
SQL (we show operator push-down in \autoref{subsec:eval:db}) and pipe primitives
for databases~\cite{Vogel:DataPipes:2023}.

\section{Prototype implementation} \label{sec:impl}

\begin{figure}[tbp]  %
    \centering
    \includegraphics[width=\linewidth]{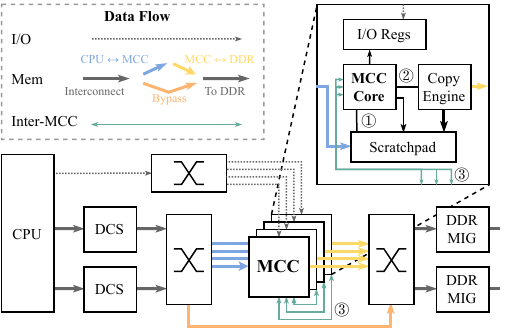}
    \caption{\system prototype}
    \label{fig:hybrics-arch}
\end{figure}

We chose to implement \system on a real hardware platform rather than
perform simulations. Simulation has great value in exploring the
parameter space for designs, but our goals in this work were
different: we wanted to first establish the practical feasibility of
the \system design, and also to identify unexpected systems issues
that arise from the abstraction that affect performance, which we
discuss in \autoref{subsec:hw-lessons}.

Our prototype uses the Enzian research
computer~\cite{Cock:Enzian:2022}, which consists of a 48-core ARMv8
\txone CPU~\cite{Cavium:ThunderX:2017} and a large FPGA in a NUMA
configuration connected by the CPU's inter-socket cache coherence
protocol~\cite{Ramdas:CCKit:2023}: the FPGA's attached DRAM
corresponds to far memory, and we implement the hardware side of
\system on this FPGA. This structure is similar to systems based on
\gls{cxl}~\cite{CXL31} or the \gls{ccix}~\cite{CCIX1}.

We use Enzian rather than existing \gls{ndp} accelerators because it
affords us greater flexibility in designing our programming model, in particular 
when it comes to the communication between the CPU and the \glspl{cp}. 
The \gls{eci} supports CXL 3.0-like\emph{symmetric coherence}, and we use it
to provide low-latency \system pipes.
We use \emph{non-coherent} far memory for the memory in our prototype, 
which models the currently-available CXL 1.1 and 2.0 hardware deployed in
production~\cite{Wang:CXLPerf:2025,Marvell:StructeraA:2024}. 

The \txone CPU has a 16MB 16-way associative \gls{llc}, and memory
transactions use fixed-size 128-byte cache lines.  The cores have no
implicit hardware prefetchers, but support explicit prefetch
hints which we use to optimize CPU-based benchmarks.

We characterize the performance of this platform below in
\autoref{subsec:basic-perf}, showing that it closely approximates the
emerging far memory accelerators in memory performance. 

\subsection{System overview}\label{subsec:overview}

\autoref{fig:hybrics-arch} shows the high-level
architecture of the \system hardware. 
\gls{eci} carries coherence messages from the CPU,
handled on the FPGA by two \glspl{dcs}~\cite{Ramdas:CCKit:2023} which form the \gls{eci}
controller, similar to the \gls{cxl} Device Coherence Engine.  
The \system implementation sits behind this infrastructure and runs at 300MHz,
using 512-bit-wide data paths throughout.

Far DDR memory is accessed via two interleaved DDR
\glspl{mig}~\cite{AMD:DDRMIG:2024}, each controlling one
32GB DDR4 DIMM.  Based on the physical address, the CPU either
accesses this memory directly or the coherence request arrives at one
\gls{mcc}.
Each \gls{mcc} comprises of a soft processor core, scratchpad memory,
the copy engine, and I/O registers for out-of-band control by the CPU
OS. 

The \textbf{core} is a 64-bit MicroBlaze-V~\cite{AMD:MicroBlazeV:2025}
in-order RISC-V soft-core wih integer multi/div, single-precision
floating-point, and bit-manipulation extensions enabled.  It runs at
300MHz in Machine mode (the highest level of privilege), and has
128KiB of fast memory for code and data.

The \textbf{scratchpad memory} holds 512 cache lines (each 128 bytes),
plus 2 64-bit time stamps per line, and can be
accessed in a single cycle from the core. It is also connected with
the ingress datapath and can be accessed by the host CPU.  

The CPU and \gls{mcc} communicate and synchronize via the scratchpad.
The \gls{mcc} can ``read lock'' or ``write lock'' a line in the
scratchpad (\circled{1} in \autoref{fig:hybrics-arch}).

A CPU \gls{llc} read of a read-locked line will not be served immediately;
instead the \gls{mcc} core is notified via a FIFO queue of messages,
indicating the line index and request type.  It can then modify the line
before releasing it to be sent to the CPU's \gls{llc}.
A CPU write to a write-locked line will store the data to the
scratchpad, but will again notify the \gls{mcc} and the write
acknowledgement will be delayed until it is unblocked by the
\gls{mcc}.

These features enable extremely efficient message-passing between CPU and
\gls{mcc} (see \autoref{subsec:msg-pass-perf} below) and are used to
implement the \system \textbf{pipe abstraction}. 

The scratchpad is also used for spawning a new \gls{cp}: the code is copied by the CPU to the scratchpad, and then
the \gls{mcc} copies code and data into private memory.

The \gls{mcc} core cannot directly access far DRAM. Instead, it instructs the
\textbf{copy engine} to transfer one or more cache lines at a time
between the scratchpad and the DDR (\circled{2} in
\autoref{fig:hybrics-arch}) by writing registers on the copy
engine. This operation is split-phase: the core continues executing
immediately, and when the operation completes, the copy engine updates
one of the line's timestamp values (depending on read or write)
with the value of a 64-bit global monotonic counter. 

The \glspl{mcc} are connected by a full mesh of
point-to-point \textbf{inter-MCC message passing pipes} (\circled{3}
in \autoref{fig:hybrics-arch}), implementing FIFOs holding 1024 words each.
In addition, each \gls{mcc} also has 32 64-bit I/O registers that are exposed to
the CPU in the I/O address space and accessible from the \gls{mcc} core with a
higher latency (9 cycles) and used for management signalling, for
example to reprogram a new \gls{cp}.

Finally, all critical data paths have \textbf{performance
counters} attached, which we use in \autoref{sec:eval} to provide insight into
application runtime behavior.

With this design, we can comfortably fit four \glspl{mcc} on 
Enzian's FPGA.   Below, we characterize the performance
of the prototype and compare with related hardware. 

\subsection{Performance characterization}

To characterize the basic performance of \system as promised in \autoref{sec:bg},
we measure its performance and compare the results
with emerging hardware reported in the
literature~\cite{Wang:CXLPerf:2025,Sun:2023:DemystifyingCXLMemory,Tang:CXLPerf:2024,WangXi:PerfCXL:2025,Berger:Octopus:2025,
  Li:Pond:2023} and product
announcements~\cite{Marvell:StructeraA:2024,GoogleCloudC4:2026,GoogleCloudC4A:2026}.

\subsubsection{Bandwidth, latency, and compute}\label{subsec:basic-perf}

We start with the memory bandwidth, access latency, and compute power
of \system.
From the host CPU, latency is measured by timing uncached reads to far
memory. We measure peak sequential read bandwidth using L1 and
L2 prefetch instructions to pick the best configuration. The
numbers are reported in \autoref{tab:basic-perf} as \emph{external}
(off-chip, through the interconnect) bandwidth and latency.
From the \gls{mcc}, memory accesses are performed by the copy engine. The time is
measured from the \gls{mcc} issuing a read/write command to the acknowledge arriving
back to the scratchpad memory. The numbers are reported as \emph{internal}
(on-chip) bandwidth and latency in \autoref{tab:basic-perf}.

To estimate the \mcc's compute power, we use single-core
Dhrystone~\cite{Weicker:Dhrystone:1984}, a simple benchmark to
accommodate the core's limited resources. We use Intel Xeon
Scalable Emerald Rapids to approximate the CPU of emerging hardware
and Google Axion to approximate the Marvell \structera NDP cores. 
Google Axion uses the same ARM Neoverse V2 cores 
as the \structera, but clocked at 3GHz (extracted from \code{dmidecode}).
We run Dhrystone on both Intel Xeon~\cite{GoogleCloudC4:2026} and
Google Axion~\cite{GoogleCloudC4A:2026} using Google Cloud VMs.

\begin{table}[tbp]  %
  \caption{Memory performance comparison}
  \small
  \centering
  \begin{tabular}{c|cc|cc}
                                                                          & \multicolumn{2}{c|}{Our prototype} & \multicolumn{2}{c}{Emerging hardware} \\
                                                                          & External         & Internal        & External         & Internal           \\ \hline
\multirow{2}{*}{\begin{tabular}[c]{@{}c@{}}Peak\\ BW\end{tabular}} & 9.1 GB/s         & 17.5 GB/s       & 30--80 GB/s      & 100--200 GB/s      \\
                                                                          & \multicolumn{2}{c|}{1:1.92}        & \multicolumn{2}{c}{1:2 to 1:3}        \\ \hline
\multirow{2}{*}{Latency}                                                  & $715$ ns    & $238$ ns    & 200--300 ns      & 80--120 ns         \\
                                                                          & \multicolumn{2}{c|}{3.00:1}        & \multicolumn{2}{c}{2:1 to 3:1}       
\end{tabular}

  \label{tab:basic-perf}
\end{table}

\begin{table}[tbp]  %
  \caption{Compute performance comparison}
  \small
  \centering
  \begin{tabular}{c|cc|cc}
                                                                & \multicolumn{2}{c|}{Our prototype} & \multicolumn{2}{c}{Emerging hardware}                                                                                                                        \\
                                                                & CPU              & MCC             & CPU                                                                         & NDP cores                                                                      \\ \hline
Frequency                                                       & 2 GHz            & 300 MHz         & \textasciitilde2.3 GHz                                                      & 3.2 GHz                                                                        \\ \hline
\begin{tabular}[c]{@{}c@{}}Single-core\\ Dhrystone\end{tabular} & $0.15\times$     & $0.03\times$    & \begin{tabular}[c]{@{}c@{}}$1\times$\\ @ \textasciitilde 2 GHz\end{tabular} & \begin{tabular}[c]{@{}c@{}}$1.25\times$\\ @ \textasciitilde 3 GHz\end{tabular} \\ \hline
\# Cores                                                        & 48               & 4               & 60+                                                                         & 16                                                                            
\end{tabular}

  \label{tab:basic-perf-cpu}
\end{table}

While our prototype has $2-3\times$ higher latency
(both internal and external) and $3-5\times$ lower bandwidth, the \emph{ratios} of external
versus internal bandwidth/latency are remarkably close to the emerging
hardware.  Moreover, the system in operation is orders of magnitude
faster than a simulator.
As we are interested in the performance \emph{gains} of offloading workloads
to the \gls{mcc}, these ratios establish the validity of results we present in
\autoref{sec:eval}.

The greatest divergence in the \system prototype is the relatively
poor \emph{compute power} of the \gls{mcc} (\autoref{tab:basic-perf-cpu}): even using all four MCC cores
optimally still had much less througput than a single \txone core, due to
the FPGA's low clock frequency.  While platforms like
UPMEM~\cite{Gomez-Luna:PrIM:2022} also have weak accelerator cores,
\structera~\cite{Marvell:StructeraA:2024} has much more compute power
close to memory.  Remarkably, as we show in \autoref{sec:eval}, the
system prototype still delivers considerable benefits across a range of
applications; scaling up compute in the \gls{mcc} would only improve
this further.

\subsubsection{Message passing using pipes} \label{subsec:msg-pass-perf}
As described in \autoref{subsec:overview}, \system implements pipes between the
CPU and \gls{mcc} using cache line transactions. To evaluate the throughput
and latency, we implement a minimal message-passing application. To show the
effectiveness of using cache lines, we also evaluate pipes
using conventional MMIO registers.

The application passes a list of 64-bit integers as messages from the CPU to a
single \gls{mcc}. To maximize the message
passing throughput, multiple integers (messages) are packed into a cache line.
The register-based pipe is similar, except that
each register can hold only one integer. Additionally, CPU reading/writing MMIO registers
cannot be blocked by the \mcc. To avoid losing messages, the
CPU needs to poll until the register is marked as clean by the \mcc. We
vary the number of CPU threads and cache lines/registers per thread. 

The \mcc \cp verifies that no messages are lost by
accumulating the first integer in the cache line/register. We
make this computation as lightweight as possible to focus on the performance of
the pipes rather than the (low) instruction throughput of the \mcc, as seen in \autoref{subsec:basic-perf}.

\begin{figure}[tbp]  %
    \centering
    \hspace{-1cm}\includegraphics[width=0.95\linewidth]{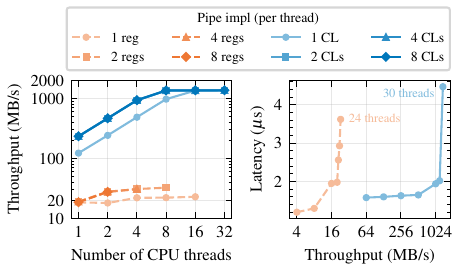}
    \hfill%
   \begin{subfigure}[t]{0.50\linewidth}
        \centering
    	
    	\caption{Throughput (log axis), CPU to one MCC}
    	\label{fig:msg-pass-tp}
    \end{subfigure}%
    \hfill%
    \begin{subfigure}[t]{0.48\linewidth}
        \centering
        \caption{Throughput vs latency, 1 reg/CL per thread}
        \label{fig:msg-pass-tp-lat}
    \end{subfigure}%
    \caption{Throughput and latency of pipes.}
    \hfill%
\end{figure}

\autoref{fig:msg-pass-tp} shows the message passing throughput to one \mcc.
Cache line-based message passing peaks out at about 1400 MB/s, lower
than the interconnect bandwidth (\autoref{subsec:basic-perf}) since the limiting
factor is the compute power of \mcc.  Receiving one
cache line at 300 MHz takes about 27.4 cycles on average, which includes notification,
unmarshaling and minimal verification.
For a single CPU core and one cache line, the core
immediately requests the cache line back for the next message
after flushing. This makes the core block
on the cache line in flight or being processed by the \mcc,
resulting in only 126 MB/s. With more cache lines, the CPU is not blocked: it
can write to other cache lines in the meantime. With 8 cache lines, the overhead
is amortized and the single CPU core reaches almost the max throughput.

In contrast, through I/O registers, the throughput saturates at 19.0 MB/s for a 
single thread and at 28.4 MB/s for multiple threads. The magnitude difference with the
cache line-based mechanism is mainly due to I/O registers being narrow, allowing
one integer at a time. Furthermore, MMIO is
not designed to transfer data and forces the CPU to serialize.

As shown in \autoref{fig:msg-pass-tp-lat}, the average round-trip latency over $2^{23}$ samples
using a single
cache line is $1580$ ns; with a single register this is $1200$ ns. MMIO
has slightly lower latency for the lowest throughput since it does not go through all cache infrastructure.
As throughput increases, however, the latency of MMIO pipes quickly degrades and becomes
worse than even the smallest message on cache line-based pipes.

This experiment suggests that message passing over cache lines is
highly efficient with low latency and high throughput. We actively use this pipe implementation in the following benchmarks.
On the other hand, while MMIO pipes have much lower throuhgput, 
their low latency for small messages is still useful for simple synchronization.

\subsubsection{\gls{cp} spawn time}\label{subsec:eval:reprogram}

Finally, we measure the average time over 50 runs to spawn a new \gls{cp} on an
\mcc{}: time between the CPU reprogramming the \mcc to receiving
achnowledgement from the \gls{cb}.
We vary the size of the \gls{cb} up to 32KiB, larger than
all \cps we use in \autoref{sec:eval}.
We use a stub \glspl{cb} padded to specific sizes with no application logic.

\begin{figure}[tbp]  %
    \centering
    \includegraphics[width=\linewidth]{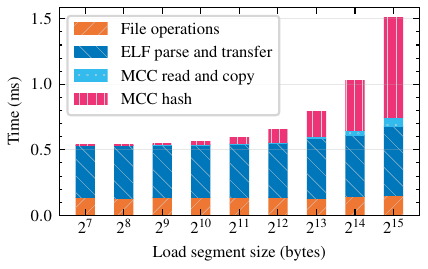}
    \caption{Spawn time}
    \label{fig:reconfig-time}
\end{figure}

\autoref{fig:reconfig-time} shows very stable spawn times (< 0.04ms SE).
For larger \glspl{cb}, the time is dominated by the \gls{mcc} verifying the binary for
correctness; this step can be omitted to optimize for peak performance.
Overall, spawning a \gls{cp} takes about 1 ms, comparable with around
500$\mu$s for small Linux process~\cite{Baumann:Fork:2019}, showing
that \glspl{mcc} can be scheduled at a fine grain, a topic we discuss
more fully in \autoref{subsec:future-os-support}. 

\subsection{Software}

Finally discuss the software components of our system.

\textbf{CPU side:} 
The CPU-side is a regular Linux application which
memory-maps the far memory, \mcc{} scratchpads and I/O registers to allow access
through regular loads and stores. As discussed in more detail in
\autoref{subsec:non-coherent-memory-access}, the lack of global coherence
for far memory and scratchpad requires the application to carefully manage the
CPU's caches using userspace cache management instructions.

\textbf{MCC side:}
The software on the MCC consists of the channel program and a support library
to control the copy engine, notifier and inter-MCC message queues, as well as
the minimum required functionality that would normally be provided by \texttt{libc}. 
Similar to the CPU side, the \cp{} uses regular loads and stores to access the
scratchpad and I/O registers. The \cp{} is compiled using the Xilinx Microblaze
compiler toolchain.

\section{Evaluation} \label{sec:eval}

Our goal in this section is partly to show that our prototype
delivers application benefits, but more importantly to show that the
\emph{\system abstractions allow efficient implementation of application
functionality}, even on the relatively low-speed \gls{ndp} cores in our
platform.  

For fair comparison with CPU-only implementation, we apply various
optimizations in the CPU baselines.   One is aggressive
application-tailored prefetching in the CPU: for example
\autoref{subsec:eval:uts} and \autoref{subsec:pagerank} we start a prefetch
as soon as an address is known. We also adopt techniques from
Grappa~\cite{Nelson:GrappaTR:2014,Nelson:GrappaATC:2015}: asynchronous
accesses via coroutines and batching messages.

\subsection{Bulk memory operations}\label{subsec:eval:bulk}

Filling or copying far memory is a natural workload to offload to
\gls{ndp}, although also the most basic one.  We implement these
operations as \cps and evaluate the benefit over a CPU-only approach.
The \cps acquire source and destination addresses via a pipe from the
host application.  Our measurements do not include CPU-side cache
management.  

The CPU baseline uses \code{libc} \code{memset} and \code{memcpy}
on far memory, and ends by flushing the destination buffer from the cache
to ensure that data hits far memory.  Since this flush is
asynchronous, it then reads to last-written cache line -- an operation
we also include in the \cp for fair comparison.
We use two \texttt{memcpy} implementation \glspl{cp}.  \texttt{memcpy}
uses a series of synchronous fetches and writes between
RAM and scratchpad, whereas \texttt{memcpy\_opt} interleaves two
split-phase memory transactions at a time. 
For all setups, we calculate throughput from the end-to-end latency of
the bulk operation. We measure 100 times and report the
median and 95\% confidence interval.

\begin{figure}[tbp]  %
    \centering
    \includegraphics[width=\linewidth]{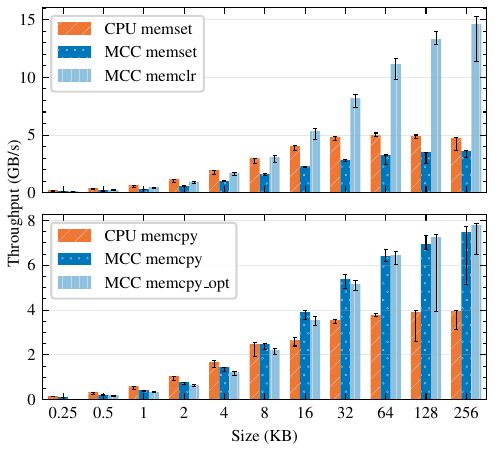}
    \caption{\code{memset} and \code{memcpy} throughput in far memory}
    \label{fig:mcc-memops}
\end{figure}

As \autoref{fig:mcc-memops} shows, a single \gls{cp} on our prototype
can still deliver superior performance when accessing more than 8KiB
of far memory. The exception is \texttt{memset}, which accepts a fill
value and 
re-initializes the cache line to write each time around its internal
loop, exposing the limits of the 300MHz core.  The \texttt{memclr}
operation (which zeros memory) shows no such issue.   Note that these
figures \emph{include the cost of starting the operation remotely from
the CPU}. 

\subsection{GUPS-like random memory access}\label{subsec:eval:gups}

This experiment evaluates the effect of performing more processing
close to memory.  We evaluate a single \mcc core's raw performance
updating random memory locations by adapting the HPCC
``RandomAccess'', which is designed to eliminate spatial and temporal
locality.  In our variant one worker generates
random indices into a large table in memory and performs 
modifications on each element.  

We implement CPU baselines with the table in local and far memory. The worker
loads the value into cache and modifies it there; we do not insert
barriers between updates, allowing the CPU to exploit pipeline
parallelism. To minimize CPU overhead due to TLB misses, the table is mapped using
2MiB large pages. 
The \cp implementation loads each entry's entire cache line into scratchpad,
modifies it, and writes it back synchronously with no parallelism.

We measure updates per second 10 times for each
table size, and report median and 95\% confidence intervals.

\begin{figure}[tbp]  %
    \centering
    \includegraphics[width=\linewidth]{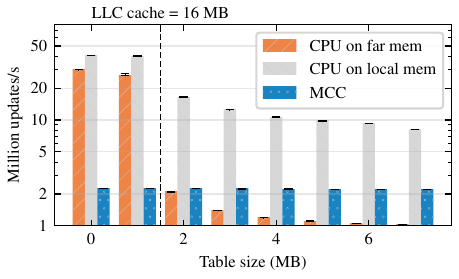}
    \caption{Table updates per second vs. working set}
    \label{fig:gpus-perf}
\end{figure}

\autoref{fig:gpus-perf} shows that when the working set fits entirely
into the CPU's L2 cache, the CPU is about 20$\times$ faster than the
\gls{cp}, but this effect disappears for far memory as soon as the
table is larger: the \gls{cp} dominates performance here, although the
CPU remains about 3$\times$ faster if it only accesses local memory.

This shows the computationally weak \mcc using \system abstractions
can outperform a CPU core when randomly accessing far memory.  A
completely synchronous \gls{cp} implementation with no parallelism can
outperform a much more performant CPU core when the workload
has little inherent locality and the working set exceeds the
CPU cache.

\subsection{In-memory database operators}\label{subsec:eval:db}

This experiment shows that memory-intensive but regular in-memory database operators 
benefit from offloading in \system by utilizing the split-phase copy engine and
the high internal bandwidth.

We implement \cps for single-table aggregation and filter operations and
compare performance with their CPU counterparts.
We generate a synthetic \textit{accounts} table with $2^{20}$ rows
with \emph{id} as primary key and a \emph{balance} column populated with 
random numbers.  Each table is an array of C structs with 32-bit \texttt{id}s and \texttt{balance}s. 
We optionally add padding inside each C struct to simulate row sizes of 16, 32, 64 and 128 bytes.

The CPU baseline iterates over the row store in a tight loop, manually 
prefetching 64 cache lines ahead.  We map the table with 2 MiB
hugepages for effective prefetching: the prefetch is a no-op on TLB miss.
The \mcc \cp prefetches table chunks into scratchpad.
The CPU invokes the \cps using I/O register-based pipes as discussed in 
\autoref{sec:impl}.

\begin{figure}[tbp]  %
    \centering
    \includegraphics[width=\linewidth]{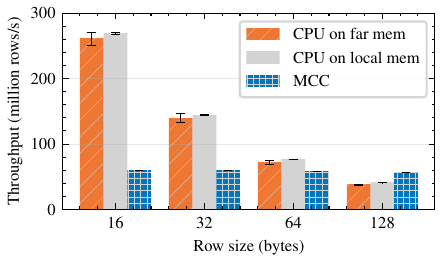}
    \caption{Single 4B column sum throughput}
    \label{fig:dbops-aggregation}
\end{figure}

\textbf{Sum of a single column:} 
\autoref{fig:dbops-aggregation} shows the throughput of a sequential scan and
sum of the \emph{balance} column for different row sizes.
The \mcc is compute bound in all cases with a stable throughput
around 57M rows/s, showing the limits of the 300MHz core.

For rows above 32 bytes in size the CPU is memory bound.  For 128
bytes and more the \gls{cp} outperforms the CPU with at least 39\% higher throughput. 
At this row size, the CPU cache can only use one 4B value from each 128B
cache line, while the \mcc explicitly prefetches a constant amount of data for
all row sizes.   For larger row sizes, a \gls{cp} is a good fit for
aggregation operators.

\textbf{Range filter:} 
We now select rows where \emph{balance} is between the parameters
\texttt{min} and \texttt{max}, and vary these to achieve different
selectivity.  Here we use 128B rows to stress make the CPU baselines
memory bound. 

The CPU baseline reads the table but does not materialize the filtered result.
The \gls{cp} materializes the results, but does not compact each row.
It returns the result rows to the CPU in two ways:
\textbf{MCC Copy} copies the filtered rows back to far memory and
coalesces copy operations of adjacent rows when possible, using the
copy engine.  This does not include transferring results to the CPU or
managing the latter's caches.
In contrast, \textbf{MCC Streamed} streams
the filtered rows back to the CPU through a \system pipe.

\begin{figure}[tbp]  %
    \centering
    \includegraphics[width=\linewidth]{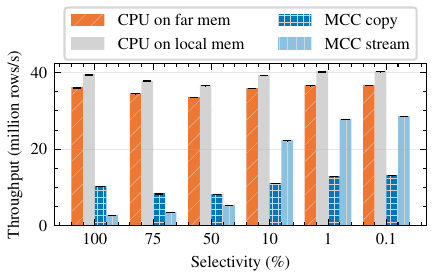}
    \caption{Range filter throughput vs. selectivity}
    \label{fig:dbops-filter-128}
\end{figure}

\autoref{fig:dbops-filter-128} shows the limits of the \gls{mcc}
processor: the \gls{mcc} is heavily compute bound and the CPU is
consistently faster is all cases. However, the MCC Streamed results
still show the benefits of reducing data transfer between host and
memory, considerably outperforming MCC Copy for many cases, and
approaching the CPU performance for high selectivity (note that MCC
Streamed results end up in the CPU's cache).

Considering the architectural disadvantage of the \mcc, we conclude that the
\mcc is suitable for offloading memory-intensive database operators with
regular access patterns.

\subsection{Unbalanced tree search}\label{subsec:eval:uts}

We now look at benchmarks which exploit features of \system --
split-phase memory operations, and low-latency pipes -- not found in
other \gls{ndp} programming models. 

\gls{uts} exhaustively searches dynamically generated unbalanced
trees and exhibits both irregular memory access patterns and
unpredictable load distributions in the different tree branches.
Our implementation on \system uses multiple
\glspl{mcc} for dynamic load balancing, as well as aggressively
parallelizing memory access on each \gls{mcc} to hide latency. 

We measure a single-\gls{mcc} implementation that uses the copy
engine to keep 8 memory read transactions in flight at a time with a 
software pipeline, allowing the \gls{cp} to fetch node metadata and
child nodes asynchronously.

We also use a 4$\times$ \gls{mcc} version of \gls{uts} using
inter-\gls{mcc} pipes to dynamically assign tree nodes to \glspl{mcc}.
Each \gls{mcc} starts traversing the tree at a given node, dispatching
child nodes equally to itself and its peers using pipes.  We use a
form of credit-based termination
detection~\cite{Mattern:Termination:1989}. 

We compare with four baselines on the CPU. One is a single threaded recursive
DFS algorithm with the tree in local memory and no prefetch hints, and
one is the same but using far memory.  The third is the best
achievable far memory DFS algorithm using Grappa-style prefetch hints
and software pipelining, and the fourth is a 4-thread, 4-core
parallel, pipelined version. 

We use the T1L, T2L, and T3L trees defined by the \gls{uts} benchmark. 
Each has about 100 million nodes. 
T1L is shallow, extremely bushy, and relatively regular. T2L is 
similar but with more irregular branching. T3L is a highly
unbalanced and deep binomial tree.

\begin{figure}[tbp]  %
    \centering
    \includegraphics[width=\linewidth]{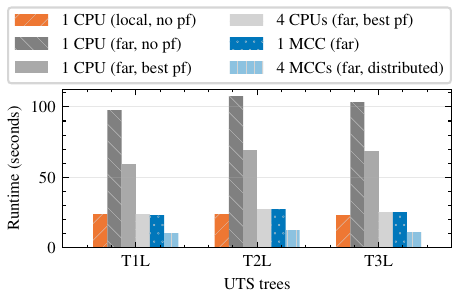}
    \caption{Unbalanced tree traversal time (pf: prefetch)}
    \label{fig:uts-perf}
\end{figure}

The results are reported in \autoref{fig:uts-perf}.  Far memory
imposes a slowdown of 4.0x-4.5x for the simple CPU version, which is
then improved by around 40\% using careful prefetching. Parallelizing
this with four CPU cores is about as fast as both unoptimized
local-memory traversal on one CPU, and traversing the tree using a
single \gls{cp}.  four-core far memory implementation.  However,
\emph{parallelizing on four \glspl{mcc}}, achieves a 2.3$\times$
throughput increase -- the best performance.

These results show that the \system copy engine allows a \gls{cp} to 
efficiently hide memory latency, the pipe abstraction 
allows multiple MCCs to effectively, dynamically parallelize irregular
workloads -- in this experiment there is only a 4\% variance in the
number of nodes processed on each \gls{mcc}. 

\subsection{PageRank} \label{subsec:pagerank}

The final application is PageRank, which demonstrates fine-grained
coordination between the CPU and \gls{cp} working on different parts
of a problem, using CPU-\mcc pipes and asynchronous memory
transfers. It serves as a complete example of accelerating real
applications with the \system abstractions, and we also show that the
\system model is flexible to support various combinations of CPU
core and \glspl{mcc}.

We evaluate our implementations on graphs listed in \autoref{tab:pr-graphs}.
\code{kron26} is a generated Kronecker
graph~\cite{leskovec2010kronecker} with
scale factor 26 that mimics a highly skewed, power-law degree
distribution. \code{twitter} and  \code{sk-2005} are real-world social
network and web crawls with the similar distribution. \code{road} and
\code{urand} are typically not used for PageRank, but reveal valuable
insights for certain design choices of \system.

The algorithm (\autoref{fig:pr-pseudo-code-diff}) consists of phases.
In sequential phases 1 and 3, the score arrays are processed
linearly. In the score accumulation phase (2), graph traversal
issues sequential accesses (the \gls{csr} format) but the score array
is accessed randomly. For CPU-only implementations, this accounts for
more than 95.4\% of run time for all implementations on all graphs in
\autoref{tab:pr-graphs} except \code{road} (min 76.3\%). 

With \system, we offload this part to the \mcc, which exploits the on-chip
latency and bandwidth to accelerate random accesses. The other sequential parts,
including the graph traversal, remain on the CPU for its strong compute power.
We evaluate both cache line-based and register-based pipes described in
\autoref{sec:impl}. Node IDs are 32-bit integer. When using the cache line-based
pipes, we use the message format of \code{(u, count, v0, v1, ...)}. Up to 30
\code{v} node IDs are packed into one cache line.

For the CPU-only baseline, we adapt the GAP benchmark~\cite{Beamer:GAP:2017} (which
performs the same pull-based PageRank as
\autoref{fig:pr-pseudo-code-diff}), and insert the optimal set of
prefetch hints. 
We compare with an implementation using 4 \glspl{mcc}. 

\begin{figure}[tbp]  %
    \hfill\begin{minipage}{0.45\textwidth}
    	\centering
    	\input{code/pagerank-cpu-diff.tex}
    \end{minipage}
    \caption{Pseudocode for PageRank. Orange indicates the CPU-only
    	    code. Blue for the collaborative CPU-MCC code.}
    \label{fig:pr-pseudo-code-diff}
\end{figure}

\begin{table}[tbp]  %
    \caption{PageRank graphs}
    \small
    \centering
    \begin{tabular}{lrrrr}
\toprule
Graph & Nodes/$10^6$ & Edges/$10^6$ & Max in-degree & CLFR \\
\midrule
\texttt{kron26} & 67.1 & 1051.9 & 1,003,753 & 0.997 \\
\texttt{twitter} & 61.6 & 1468.4 & 770,155 & 0.927 \\
\texttt{sk-2005} & 50.6 & 1930.3 & 8,563,807 & 0.125 \\
\texttt{road} & 23.9 & 57.7 & 9 & 0.848 \\
\texttt{urand25} & 33.6 & 1073.7 & 113 & 1.000 \\
\bottomrule
\end{tabular}

    \label{tab:pr-graphs}
\end{table}

\begin{figure}[tbp]  %
    \centering
    \includegraphics[width=\linewidth]{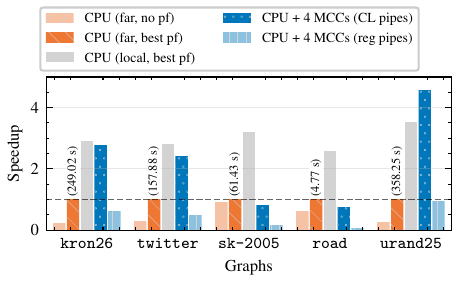}
    \caption{PageRank speedups, one CPU core}
    \label{fig:pr-perf}
\end{figure}
\begin{figure}[tbp]  %
    \centering
    \includegraphics[width=\linewidth]{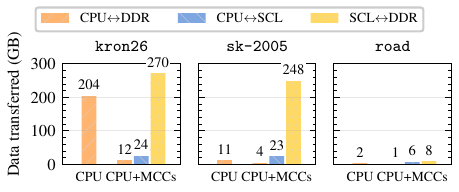}
    \caption{Data transferred in PageRank}
    \label{fig:pr-stats}
\end{figure}

\autoref{fig:pr-perf} shows relative speedup against 
the far memory baseline with optimal prefetching.  CPU and local memory is
2.5-2.8$\times$ faster than the baseline with no prefetch. The \system
implementation with cache line-based pipes is 0.7--4.5$\times$ faster than the
baseline, and this achieves 4.6--14.6$\times$
performance than the register-based counterpart.

On \code{kron26} and \code{twitter}, \system achieves speedups of 2.77$\times$
and 2.40$\times$ respectively, mainly from the \mccs handling
read-intensive score gathering.  This is confirmed by \system
performance counters (\autoref{fig:pr-stats}): for \code{kron26}, most reads terminate
on-chip with 5.66$\times$ less data transferred. However, as the \mccs
have no cache, DDR traffic increases by $38\%$.

\code{sk-2005} is also highly skewed, but since web crawls are
lexicographically sorted, node IDs connected to a common node are
highly clustered, making the score update very cache-efficient, as
\autoref{tab:pr-graphs} shows: the \gls{clfr} of
\code{sk-2005} is significantly lower. Indeed, the CPU only generated
11 GB traffic to the far memory. In comparison, the cacheless
\gls{mcc} incurs a 23.46$\times$ DDR access amplification. 
While the traffic incurred by message passing remains similar as \code{kron26}
(23 GB), it is now 2.51$\times$ higher than CPU performing the score gathering,
resulting in \system being 22.5\% slower than the baseline.  This implies that
\mccs can still benefit from a cache for certain workloads.

For graphs not traditionally used with PageRank, the results reveal critical
characteristics of \system. \code{road} is derived from a road network with much
lower degrees. The current message format of \system packs at most 30 neighbors
of \emph{one node} into a cache line, resulting in under-utilitized cache lines
for this graph.  As a result, the message (un)marshal overhead dominates, making
\system slower than the baseline. \autoref{fig:pr-stats} also confirms a
6$\times$ increase in \gls{eci} traffic.  This leads to a critical insight: the
message format should be tailored to application.  We discuss how this motivates
the need of a high-level programming model and OS support in
\autoref{subsec:future-os-support}. On the other hand, \code{urand}, being
uniformly random with 1.0 \gls{clfr}, is the worst possible case for the CPU. On
this graph, \system is $4.56\times$ faster than the baseline, even outperforming
the base case of CPU on local memory.

Overall, we conclude that when the CPU is memory-bound, \system mitigates the
memory bottleneck and saves data movement over the interconnect.

\begin{figure}[tbp]  %
    \centering
    \includegraphics[width=0.75\linewidth]{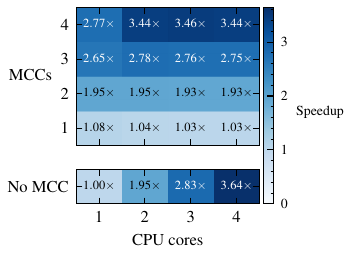}
    \caption{\code{kron26} speedup with varying numbers of CPU cores and MCCs}
    \label{fig:pr-scaling}
\end{figure}

Furthermore, as discussed in \autoref{sec:design}, \system allows flexible scheduling
betweeen CPU cores and \mccs. To demonstrate that, we execute
\code{kron26} with varying numbers of CPU cores and \mccs.
\autoref{fig:pr-scaling} shows the normalized speedups.

When a CPU core is paired with only one \mcc, the execution is $8\%$ faster.
With more \mccs, the speedup increases. 
However, using 4 \mccs is only marginally faster than using 3. 
If we increase the number of CPU cores (moving to the right on the row), the
performance improvement stagnates.

When varying the number of CPU and MCC cores, we found that  
a CPU core with only one MCC is initially bounded by the low compute power of
the \mcc, but once we reach 4 \mccs per core, the CPU becomes the bottleneck as
it cannot issue messages fast enough. Adding more CPUs further improve the
performance. But as the ratio of \mccs \emph{per CPU} decreases (moving to the
right of the row), the \mcc again becomes a bottleneck.

With this, we conclude that \textbf{CPUs
and \mccs can be flexible combined}. \system is neither contrained by a fixed
mapping nor requiring a power-of-two numbers of units, unlike previous \gls{pim}
solutions. This also paves the way to our future work of an end-to-end
virtualizable and managed system (\autoref{subsec:future-os-support}).

\section{Discussion} \label{sec:discussion}

By building a real hardware prototype, we gained valuable knowledge about
the interplay of system components in a far-memory accelerator.

\textbf{Memory architecture of \ndp units:}
A regular CPU accesses memory with load/store semantics and uses caches and
prefetchers to accelerate this, while \system's abstractions limit
loads and stores on the \gls{mcc} to private and scratchpad memory,
and instead enables explicit data movement to the far memory via the
split-phase copy engine.
We argue that \ndp processors benefit from having more explicit and
deterministic data movement to and from DRAM, shown by the
applications in \autoref{sec:eval}, and that these benefits would
still exist with more powerful cores on the \gls{mcc}. 

As \autoref{subsec:pagerank} shows, some applications still
benefit from a hardware-managed cache to reduce DRAM traffic, and
emulating this using a larger scratchpad would incur additional
\gls{mcc} software overhead.  Adding a cache to the \ndp between the
copy engine and DRAM would provide additional, automatic hardware
management of fast memory.

\textbf{Synchronization mechanisms.}
\autoref{subsec:pagerank} shows that efficient synchronization
between the CPU and the \mcc is essential to performance gain for a
significant class of applications. 
We argue for efficient hardware synchronization mechanisms for
high-performance \ndp systems. 
Our implementation's hardware notifier removes the need to poll for requests or responses both on
the \mcc and the CPU, significantly reducing the overhead on both sides.

\textbf{CPU architecture for far memory.}
\system's abstractions work with any standard CXL-compatible CPU and
\textbf{do not} require any customization on the CPU side.
However, we notice that architectural parameters of the CPU affect the
\emph{performance} of the far-memory extended system.
For example, request buffers on CPU are traditionally sized to saturate the
bandwidth given the local DDR latency. With increased latency of the far memory,
it can become the bottleneck.

CPU caches and prefetchers also play critical roles.  Programming in
the lack of cache coherence requires frequent cache flushes. The
implementation of the flush instructions affects the performance, as observed by previous work~\cite{Wang:CXLPerf:2025}.
Hardware prefetchers potentially hide far-memory access latency, but with
non-coherent memory, it can break our synchronization mechanism and
prefetch stale data.

\subsection{Programming without cache coherence}
\label{subsec:non-coherent-memory-access}

While the CXL standard introduces symmetric coherence in version
3~\cite{CXL31} and version 4~\cite{Benavides:IntroCXL4:2025}, existing
and emerging CXL hardware implements CXL 1.1 and 2.0 without hardware
cache coherence. The Enzian \gls{eci} protocol, based on a NUMA
interconnect, naturally supports symmetric coherence with
back-invalidation~\cite{Cock:Enzian:2022,Ramdas:CCKit:2023}. However,
we chose to implement \system using non-coherent far memory and
scratchpad memory to model emerging CXL hardware.

Programming without cache coherence imposes challenges to application
developers. First, written data can reside in the cache as stale cache
lines. To make it visible to \mcc (and by extension, other nodes
connected to the CXL memory pool), the CPU needs to flush cache lines
explicitly (required by the CXL standard~\cite{CXL31}).

On the other side, writebacks to far memory are silent. Nodes connected to
far memory must synchronize accurately to avoid reading stale
data or half-written cache lines. Strong and weak memory models
further complicate this, and application developers need to place memory
barriers in correct locations and signal other nodes at correct time. 

As discussed in \autoref{subsec:overview}, \system allows a \gls{cp}
to lock cache lines in the scratchpad memory and get notified on CPU
reading or writing. This novel synchronization mechanism not only
enables efficient pipes, but also help developers reason about data
races due to the lack of cache coherence. For
example, it is guaranteed that data in an already locked cache line never
changes due to concurrent CPU writebacks, until the \mcc unlocks it.
Nevertheless, this mechanism is limited to the scratchpad (few cache
lines to track) but not applicable to the whole far
memory. Application developers are still required to be aware of the
exact data placement and coordinate the CPU and the \mcc, and if they
are used to coherent memory may not expect data races \emph{within} a
cache line. In hardware, writebacks happen at cache line 
granularity, and with coherence maintaining exclusive ownership, racing
writes to the same line (even if the changed bytes do not overlap) result
in lost writes.

Even though we were aware of this issue, we still ran afoul
of it when debugging PageRank and the DB filter operator on \system.  
Note that these challenges do not originate from \system but from
non-coherent far memory systems in general.
Raising the abstraction level of \system to further assist programmers
in managing non-coherent shared memory is an important direction. 

\subsection{Can LLMs use the programming model?}

A question today is whether LLM coding agents can
effectively use a new programming model without prior training. 

We manually implemented a message-passing integer accumulator
\cp using the \system's write-lock C API, correctly managing the
cache. We then requested Gemini 3.1 Pro~\cite{GeminiArxiv2025}
to reproduce this accumulator \cp given a description of \system, API
documentation and a sample application exercising both the CPU and
\mcc. We emphasized the need for cache management.

Gemini produced code that \textit{uses the write-lock C API correctly}
but did \emph{not} handle half-written cache line writebacks. We
then indicated that the program fails when the system is \emph{under
load}. Gemini managed to identify ``[u]nintended, hardware-initiated
cache evictions''.  The proposed solution, however, was to add a
sequence number in each cache line; the updated code breaks alignment
and misses a barrier.  In contrast, the ``correct'' \system solution
combines a flag with the write-locking mechanism, reliably sustaining maximum
throughput.  A second attempt ended in Gemini ``thinking'' for more than 
2.5 minutes and ending up distorting the
application's intension; we terminated the session at this point. The
full chat history and generated code are clearly marked and made
available in our artifacts.

While not a comprehensive evaluation of LLM capabilities, this small
experiment highlights the \emph{challenge of cache management} for
\system but also for emerging far memory systems in general,
motivating future work towards usable high-level abstractions that
hide the hardware nuances from application developers or coding
agents.

\subsection{Further compiler support}
\label{subsec:future-os-support}

The \system abstractions are applicable to a range of far memory
accelerators, but remain fairly low-level. 
The next step is to raise the abstraction by
integrating specialized
compiler support for our programming model.
Recent work such as Triton~\cite{Tillet:Triton:2019} shows the
feasibility of efficient automatic prefetching and scratchpad memory
management in a compiler, and can be applied to further abstracting
\system's split-phase copy engine.

Such compiler support would also enable more efficient sharing of
\gls{mcc} hardware. At present, each \gls{cp} has exclusive use of a
\gls{mcc}, but should be feasible to compile multiple \gls{cp} into a
single binary which interleaves accesses by different application code
according to static schedule chosen by the compiler, as in some hard
real-time systems~\cite{kopetz}.

\bibliographystyle{ACM-Reference-Format}
\bibliography{references}

\end{document}